\documentclass[preprint,12pt]{elsarticle}
\usepackage{graphicx}
\usepackage{hyperref}
\usepackage{float}
\usepackage{amssymb}

\begin{document}
\begin{frontmatter}

\title{First Principles Calculation of Field Emission from Nanostructures using Time-Dependent Density Functional Theory: a Simplified Approach}

\author{Sherif Tawfik, Salah El Sheikh and Noha Salem}
\date{\today}

\begin{abstract}
We introduce a new simplified method for computing the electron field emission current in short carbon nanotubes using ab-initio computation in periodic simulation cells. We computed the evolution of the wave functions using Time-Dependent Density Functional Theory, where we have utilized the Crank-Nicholson propagator. We found that in pristine carbon nanotubes, the emitted charge tends to emerge mostly from electrons that are concentrated at the nanotube tip region. The charge beam concentrates into specific channel structures, showing the utility of carbon nanotubes in precision emission applications.
\end{abstract}

\end{frontmatter}

\section{Introduction}

Field emission, the escape of electrons from the surface into vacuum under the influence of an applied electric field, has been receiving research attention owing to its theoretical, as well as commercial, significance \cite{TOhwaki}. The literature on field emission form various nanostructures is enormous. Unique field emission properties were discovered in carbon nanotubes owing to their high aspect ratio and mechanical and chemical stability, making them the best candidate for flat-panel displays and electron microscopy probes \cite{YoshikazuNakayama,YSChoi,MMeyyappan}. Field emission from graphene was also studied \cite{ZhongShuai}. Field emission was also investigated in nanodiamonds \cite{CWangBZhengWTZheng}, buckyball coating \cite{Tumareva,Tumareva2,Toshiki} and a variety of nanoclusters \cite{SZDeng,GNFursey}. Among the number of nanostructures investigated for field emission, carbon nanotubes proved, both theoretically and experimentally, to possess strongest field emission characteristics.

Field emission was understood as a realization of quantum tunneling through the simple Fowler Nordheim theory that treats a one dimensional system composed of a free electron and a trivial triangular potential barrier, thus without any consideration for scattering effects \cite{fn}. The escape of the electron through the potential barrier can be modeled using the WKB approximation, yielding the following relationship between the extracted field emission current and applied electric field

\[
I=6.2 \times 10^{-6} \frac{\mu^{1/2}}{(\chi+\mu)\chi^{1/2}}F^{2}e^{-2 \times 10^{8}\chi^{5/2}/F},
\]

\noindent where $\chi$ is the thermionic work function, $\mu$ is the familiar parameter in the electron distribution in the Fermi-Dirac distribution function.  However, at the nanoscale, approximating the nanotip using the sharp potential well is far from accurate \cite{SeungwuHan2}. Thus was the need for \textit{ab-initio} computational methods to simulate field emission at the scale of the atom.

In recent years, field emission was simulated using a variety of \textit{ab-initio} computational schemes, including time-dependent density functional theory (TDDFT) \cite{SeungwuHan2,SeungwuHan3,SeungwuHan,MasaakiAraidai}, Kubo formalism \cite{JiLuo}, transfer matrix formalism \cite{AMayer3} and Landauer-Buttiker formalism \cite{TOhwaki}. Here we present an adaptation of TDDFT. In Sec. \ref{sec:computationaldetails} we provide a theoretical overview of our computational scheme, which is based on the implementation of the Octopus code \cite{Octopus}. We also indicate where we have altered the original code in order to compute the charge in a portion of the simulation box.

\section{Computational Details}
\label{sec:computationaldetails}

\subsection{Ground State Calculation}

We performed Density Functional Theory computation using the Local Density Approximation for exchange and correlation potential by Pedrew and Zunger \cite{PedrewZunger}. States are expanded in terms of plane waves,

\[
\psi_{n}(\textbf{r})= \sum_{\textbf{G}}\psi_{n}(\textbf{G})e^{i\textbf{G}\cdot\textbf{r}},
\]

\[
\psi_{n}(\textbf{G})=\frac{1}{V}\int d\textbf{r}\psi_{n}(\textbf{r})e^{i\textbf{G}\cdot\textbf{r}},
\]

\noindent where $V$ is the super-cell volume, and $\psi_{n}(\textbf{r})$ and $\psi_{n}(\textbf{G})$ are related by a three-dimensional Fourier transform. Atomic orbitals are substituted with Troullier-Martins pseudopotentials. The density matrix is mixed according to the Broyden mixing scheme \cite{Broyden}. We perform the ground state calculation in the absence of the external electric field, and turn on the field during the time-dependent calculation.

\subsection{Time-Dependent Density Functional Theory}

\par The solution of the time-dependent Kohn-Sham equations

\[
i \frac{\partial}{\partial t}\psi_i(r,t) = \left[ -\frac{\nabla^2}{2} + v_{\rm KS}(r,t) \right]\psi_i(r,t) 
\]

\noindent has the following exact expression:
\[
\psi_j(T) = \mathcal{T}\!\!\exp\left\{ -i\!\!\int_0^{T}d\tau H(\tau)\right\} \psi_j^{(0)}\,\!,
\]

\noindent where $\mathcal{T}\!\!\exp\,\!$ is the time-ordered exponential, which is a short-hand for:

\[
\psi_j(T) = \left\{\sum_{n=0}^{\infty} \frac{\left(-i\right)^n}{n!} \int_0^t d\tau_1 \cdots \int_0^t d\tau_n H(\tau_1) \cdots H(\tau_n)\right\}\psi_j^{(0)}\,\!
\]

If the Hamiltonian commutes with itself at different times, we can drop the time-ordering product, and leave a simple exponential. If the Hamiltonian is time-independent, which makes it trivially self commuting, the solution is simply written as:

\[
\psi_j(T) = \exp\left\{ -iTH\right\} \psi_j^{(0)}\,\!.
\]

\noindent Unfortunately, this is not the case for TDDFT when the system is exposed to external time-dependent perturbations like electric and magnetic fields or pulsed lasers. But even without an external time-dependency, there remains the intrinsic time-dependency of the Kohn-Sham Hamiltonian, which is built "self-consistently" from the varying electronic density.

\noindent The first step to tackle this problem is to split the propagation of the long interval $[0, T]\,\!$ into $N\,\!$ smaller steps by utilizing the group-theoretic property

\[
U(T, t) = U(T, t')U(t', t)\,\!
\]

\noindent of the time evolution operator. This yields the following time discretization:

\[
U(T, 0) = \prod_{i=0}^{N-1}U(t_i+\Delta t, t_i)\,\!,
\]

\noindent where $t_0=0\,\!$, $t_N=T\,\!$, $\Delta t = T/N\,\!$. So at each time step we are dealing with the problem of performing the short-time propagation:

\[
\psi_j(t+\Delta t) = U(t+\Delta t, t)\psi_j(t) = \mathcal{T}\!\!\exp\left\{ -i\int_{t}^{t+\Delta t}d\tau H(\tau)\right\} \psi_j(t)\,\!.
\]

\subsection{Propagation Method}

\par There is a wide range of methods for computing the propagator for the time-dependent Kohn-Sham equations, and research in this area continues to grasp more attention \cite{AlbertoCastro}. These propagators solve the problem of approximating the orbitals $\psi_j(t+\Delta t)\,\!$ by the knowledge of $\psi_j(\tau)\,\!$ and $H(\tau)\,\!$ for $0\le\tau\le t\,\!$. Some methods require the knowledge of the Hamiltonian at some points $\tau\,\!$ in time between $t\,\!$ and $t+\Delta t\,\!$. We choose to utilize the Crank-Nicholson method, also known as the implicit midpoint rule. This method is the most famous propagation method used to compute the time-evolution of the Schr\"{o}dinger equation, and is based on the following implicit midpoint rule \cite{ChristianLubich,HeikoAppel,AlbertoCastro}, which is the average of the forward Euler and backward Euler integration methods:

\[
i\frac{\psi_{n+1}-\psi_{n}}{\Delta t}=H(t_{n+1/2})\frac{1}{2}\left(\psi_{n+1}+\psi_{n}\right),
\]

\noindent where

\[
t_{n+1/2}\equiv 1/2\left(t_{n+1}+t_{n}\right), t_{n}=n\Delta t.
\]

\noindent Then, by straightforward algebra we obtain the following:

\[
\psi_{n+1}=\frac{1-i\Delta t\frac{1}{2}H(t_{n+1/2})}{1+i\Delta t\frac{1}{2}H(t_{n+1/2})}\psi_{n}.
\]

\noindent which is also known as the Cayley approximation of the exponential to the second order of $\Delta t$. To the third order of $\Delta t$, the Cayley approximation is

\[
\psi_{n+1}=\left[\frac{1-i\Delta t\frac{1}{2}H(t_{n+1/2})}{1+i\Delta t\frac{1}{2}H(t_{n+1/2})}+O(\Delta t^{3})\right]\psi_{n}.
\]
\noindent The problem of evaluating the evolution operator becomes that of solving the following linear matrix equation:

\[
\left[\hat{I}+i\frac{\Delta t}{2}H(t_{n+1/2})\right]\psi(n+1)=\left[\hat{I}-i\frac{\Delta t}{2}H(t_{n+1/2})\right]\psi_{n}.
\]

\par This method has two important properties: it preserves the norm of the wave function at all times (unitarity property), and exchanging $n \leftrightarrow n+1$ gives the same numerical results (time reversibility property). However, there is a tendency to prefer the other method, which is the Split-Operator method where the exponential of the Hamiltonian is split into exponentials of its composite terms (the kinetic and potential energies) \cite{AlbertoCastro}. Nevertheless, good accuracy can then be achieved by the Crank-Nicholson method only for very small time steps (such as the time step used here, which is 0.00242 fs, as the computational error is proportional to the cube of the time step at each step of the algorithm \cite{Bayfield,ChristianLubich}). A shortcoming in the Octopus code we are using is that the code uses the retarded hamiltonian $H(t_{n})$ in place of $H(t_{n+1/2})$ in 

\[
\left[\hat{I}+i\frac{\Delta t}{2}H(t_{n+1/2})\right]\psi(n+1)=\left[\hat{I}-i\frac{\Delta t}{2}H(t_{n+1/2})\right]\psi_{n},
\]

\noindent which is essentially the predictor step's equation:

\[
\left[\hat{I}+i\frac{\Delta t}{2}H(t_{n})\right]\psi(n+1)=\left[\hat{I}-i\frac{\Delta t}{2}H(t_{n})\right]\psi_{n}.
\]

\noindent The consequence of using the retarded Hamiltonian is the continuous decrease in energy every time-step because we will be using a retarded potential that arises from the retarded Hamiltonian.

\subsection{Current Calculation}

\par The amount of the electron remaining in the emitter region at time $t$ for a particular wave function is given by

\begin{equation}
\label{eq:Q}
Q_{n}(t)=\int^{z_{0}}_{0}\int\int\left|\psi\right|^{2}d\textbf{x},
\end{equation}

\noindent and in terms of the life time $\tau_{n}$ of state $n$,

\begin{equation}
\label{eq:Qfinal}
Q_{n}(t)=e^{-\frac{t}{\tau_{n}}},
\end{equation}

\noindent which implies linear behavior of $Q_{n}(t)$ in the short time interval. From this formula, the current generated by a wave function is given by

\begin{equation}
\label{eq:I}
I_{n}=e\frac{dQ_{n}(t)}{dt}\approx-\frac{1}{\tau_{n}}
\end{equation}

\noindent in the short time range. The total current is

\[
I=e\sum f_{n}\frac{dQ_{n}(t)}{dt}
\]

\noindent In order to compute the charge remaining in the lower portion of the simulation box, we integrate that lower portion each time-step for each wave function. At the beginning of the simulation, the charge is typically 1 (which works as a check for normalization consistency of the propagator used). Charge starts to decrease as time elapses, until it reaches a minimum point after which it starts increasing (due to reflection of the wave back from the upper surface of the simulation box). The integration is performed by simply summing the product of unit volume boxes of the mesh used in Octpus within the region concerned.

\section{Application to short pristine carbon nanotubes}

Field emission from carbon nanotubes has been extensively studied from theoretical and experimental perspectives, owing to their unique emission properties. Theoretical \textit{ab-initio} computation of field emission from carbon nanotubes has investigated a number of considerations that we would like to highlight from the literature. Han \textit{et al.} \cite{SeungwuHan2} stated that using a Heaviside external field results in unwanted oscillations in the charge evolution, and thus they prefered to use a constant field (converging the ground state with an external electric field) and then correct the wave functions by using a prescribed method. Screening, the damping of electric fields caused by the presence of mobile charge carriers (neighboring carbon nanotubes in a nanotube bundle), were studied by Chen \textit{et al.} \cite{GuihuaChen}. They discovered an interplay between the spacing between CNTs and their electronic structure. When the array spacing between neighboring nanotubes is three times the nanotube length, the applied external field is strongly screened. Most simulations in the literature were performed on 2D periodic systems (an infinite slab structure) with infinitely many neighboring nanotubes. Charging the nanotube (adding extra electrons) was investigated by Lou \textit{et al.} \cite{JiLuo} to simulate the realistic situation, who have also performed the simulation on isolated pristine nanotubes with 50-60 carbon atoms. Although the results reported from each \textit{ab-initio} simulation methods would give different results for the extracted current, they agree that field emission from carbon nanotubes arise mainly from states right below the Fermi level, not from the metallic continuum as in usual metallic emitters \cite{JeanMarcBonard2}.

To demonstrate our method, we studied field emission from a short (5,5) carbon nanotube, which is a metallic conductor. Our nanotube is composed of 70 carbon atoms, where the bottom carbon atoms are passivated with hydrogen atoms in order to avoid the presence of unnecessary dangling bonds. The hydrogen-passivated nanotube is $8.6 \rm\AA$ long, with a diameter of $6.75 \rm\AA$. The simulation box size is $30 \rm\AA \times 30 \AA \times 30 \AA$. The tip of the nanotube lies at the center of the simulation box, and we calculate the charge in the lower 63\% of the simulation box (which includes the lower 50\% plus an extra volume past the nanotube tip).

We adopted the time step of 0.1 au = 0.00242 fs, the same one used in  \cite{SeungwuHan2}. We also experimented with a smaller time step (0.000242 fs) and we obtained the same results, which shows that this time step is optimal for the Crank-Nicholson algorithm. In Tab. \ref{tab:pristinecarbonnanotubeperiodic} below, we show comparison between our results and those reported in Ref. \cite{SeungwuHan2} and Ref. \cite{SeungwuHan} (the cited quantities are extrapolated from graphs included in the references).

\begin{figure}
	
	\centering
\includegraphics[width=100mm]{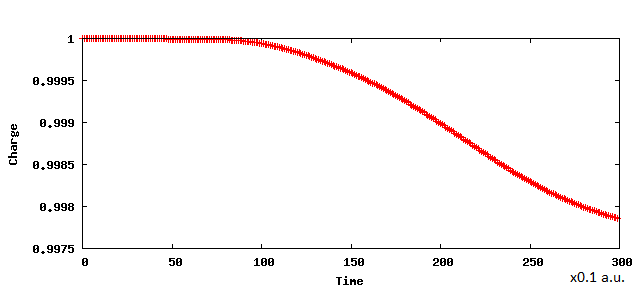}
\caption{Charge evolution for the highest energy level upon applying a field of 0.8V/$\rm \AA$. Charge is computed as the integration of the modulus of the respective wave function within the lower region of the nanotube, and units are arbitrary.}
\end{figure}

\begin{center}
\begin{table}
	\caption{Extracted current ($\mu \rm A$) from pristine carbon nanotube against an applied electric field ($\rm V/\AA$). Our results are almost close to those reported in \cite{SeungwuHan2} when the applied field is in the range 0.2-0.8$\rm V/\AA$. For larger values, reported results for the current differ from the result reported in \cite{SeungwuHan2} due to the difference in length of the nanotube used.}
	\label{tab:pristinecarbonnanotubeperiodic}
	\begin{tabular}{| p{3cm} | c | c | c | c | c | c  | c |}
		\hline
		& $0.2$ & $0.25$ &  $0.4$ & $0.6$ & $0.8$ & $1.0$ & $1.2$ \\
		\hline
		Our calculations & 0.0161 & 0.02957 &  0.1148 & 0.3958 & 0.9591  & 1.9608 &  3.8304 \\
		\hline
		Han \textit{et al.} \cite{SeungwuHan2} & 0 & 0 & 0 & 0.1 & 0.9 & 5.8 & \\
		\hline
		Han \textit{et al.} \cite{SeungwuHan} & 0 & 0 & 0 & 0.1 & 0.9 & 4.0 & 9.0 \\
		\hline
		Mayer \textit{et al.} \cite{AMayer1} &  & 0.0298 & &  & & & \\
		\hline
	\end{tabular}
	\end{table}
\end{center}

Our simulation results in a periodic simulation box are close to the results reported by Han \textit{et al.} \cite{SeungwuHan2} for the applied field in the range 0.2$\rm V/\AA$ to 0.6$\rm V/\AA$, and are very close to the result reported by Mayer \textit{et al.} \cite{AMayer1} at a field of 0.25$\rm V/\AA$ (note that Han \textit{et al.} have reported different results in two different publications,  \cite{SeungwuHan2} and  \cite{SeungwuHan} as is indicated in the Tab. \ref{tab:pristinecarbonnanotubeperiodic} above). At an applied field of $\rm 1.0 V/\AA$, both results reported by Han \textit{et al.} are slightly higher than ours, which is due to the difference in size between the nanotubes used; that is, shorter nanotubes generally yield lower current than longer nanotubes \cite{SeungwuHan3}.

\par In Fig. \ref{fig:145upper}) below, we present the evolution of the cross section of the highest energy wave functions in a pristine nanotube under a external field of $\rm 0.8 V/\AA$.

\begin{figure}
	
	\centering
\includegraphics[width=100mm]{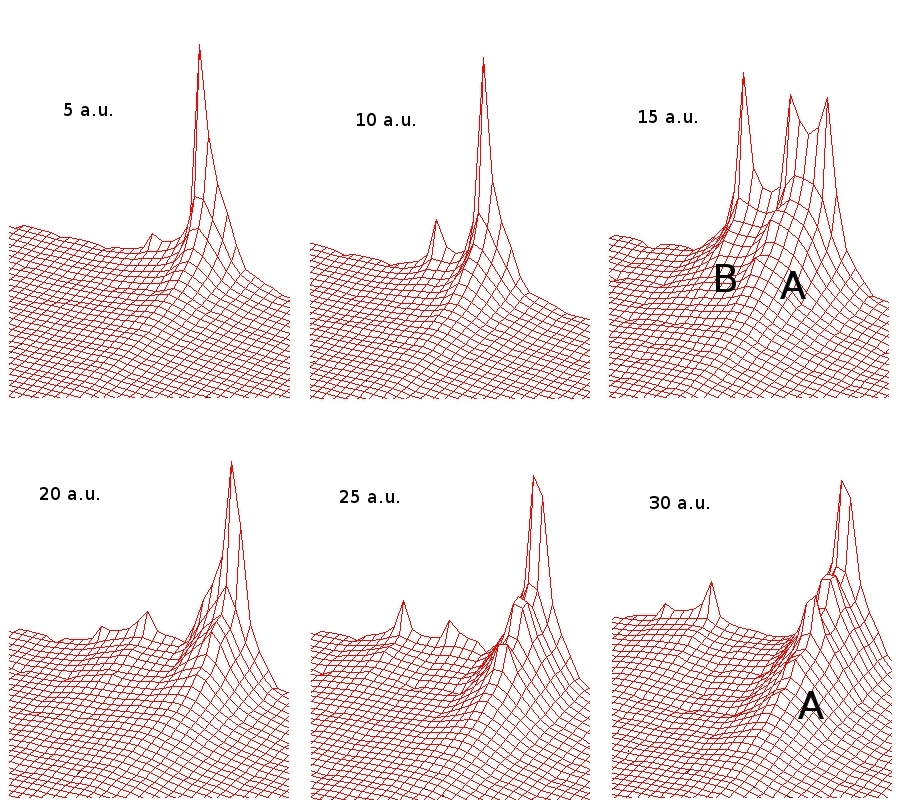}
\caption{Evolution of the wave function into the vacuum region: highest energy wave function. Units are arbitrary.}
\label{fig:145upper}
\end{figure}

\par We notice in Fig. \ref{fig:145upper} the evolution of charge of the wave function as the base of shape (A) extends forward. We also note that by 15 a.u., a branch emerges (B) from the shape, which vanished completely by 30 a.u. as the wave function becomes more concentrated into the shape (A). Also note that the relative height of shape (A) is less at time 30 a.u. than that at time 5 a.u., which is due to the fact that a portion of the charge of that wave function flows into the vacuum region. We can gain a better insight by noticing the evolution of the terminal curves at the end of the simulation box below.

\begin{figure}[H]
	
	\centering
\includegraphics[width=100mm]{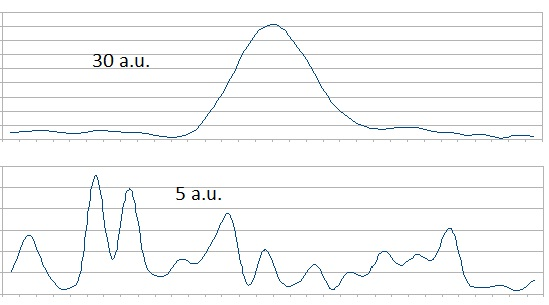}
\caption{Evolution of the wave function into the vacuum region: highest energy wave function, close to the upper surface of the simulation box. The horizontal axis is the intersection between the plane crossing the nanotube vertically and the upper face of the simulation box. The vertical axis shows the charge computed in Eq. \ref{eq:Qfinal} above, in arbitrary units.}
\label{fig:145terminal}
\end{figure}

Fig. \ref{fig:145terminal} shows clearly that the wave functions are concentrated into certain ``channels'' as time elapses, which indicates that the emitted current is concentrated towards the anode, instead of being dispersed into space. Studying such behaviors can be instrumental when the exact shape of the electron beam is concerned (when producing very high precision beams in atomic force microscopes is required). It is also of great importance whether we can influence the shape of the beam by adding certain dopants, or by performing structural modifications. This issue will be treated in a later publication.

\par In such a short nanotube, dangling bonds at the terminal carbon atoms (opposite to the tip) play a significant role. We observed that field emission in the non-hydrogen-passivated isolated carbon nanotube is almost 50\% higher than that of the hydrogen-passivated one. This is in line with the idea established in \cite{MasaakiAraidai}, that dangling bonds (arising from vacancy defects and non-passivation with hydrogen at the terminal bonds) are the prime contributors to the field emission current.

\section{Conclusions}

\par Using time-dependent density functional theory, we studied the mechanism of field emission from a short carbon nanotube. We utilized the classical Crank-Nicholson method to compute the wave function propagator, and used a small time step in order to reduce the computational error. Dangling bonds tend to considerably increase the field emission current in nanotubes; $\pi$ bonds do not contribute to field emission as much as $\sigma$ bonds. We found that the higher electronic density at the nanotube tip (which takes place due to the pentagonal structure, and which occupy higher molecular orbitals) are the prime contributors to field emission, unlike the case of metallic tips where emission is produced by the continuum of electrons in the metal \cite{JeanMarcBonard2}. We reproduced the results reported by Han \textit{et al.} \cite{SeungwuHan2} by using periodic boundary conditions. The charge cloud emerging from the tip appears to concentrate into narrower channels as the cloud approaches the opposite end of the simulation box.

\bibliographystyle{elsarticle/elsarticle-num}
\bibliography{mybib}

\end{document}